\documentclass[twocolumn,superscriptaddress,amsmath, amssymb, amsfonts,preprintnumbers,aps,prd,longbibliography,nofootinbib]{revtex4-1}

\renewcommand{\sec}[1]{\textit{#1. --- }}

\usepackage{graphicx}
\usepackage{dcolumn}
\usepackage{bm}
 \usepackage{amstext}
 \usepackage{amssymb}
 \usepackage{amsmath}
 \usepackage{graphicx}
 \usepackage{color}
 \usepackage{bbold}
 \usepackage{delimset} 
 \usepackage[caption=false,justification=justified]{subfig}
\usepackage[colorlinks=true]{hyperref}

\usepackage{tikz}
\usetikzlibrary{decorations.pathmorphing}
\usetikzlibrary{decorations.markings}
\usetikzlibrary{positioning, shapes, snakes, arrows}

\tikzset{
	graviton/.style={line width=.8pt, -latex,decorate, decoration={snake, segment length=4pt,amplitude=1.8pt, pre length=.1cm, post length=.25cm}},
	worldline/.style={gray, line width=1pt},
	worldlineBold/.style={black, line width=.6pt},
	zUndirected/.style={line width=1pt},
	zParticle/.style={line width=1pt,postaction={decorate},decoration={markings,mark=at position .6 with {\arrow[#1]{latex}}}},
	zParticleF/.style={line width=1pt,postaction={decorate}},
	cscalar/.style={line width=1pt,postaction={decorate},decoration={markings,mark=at position .6 with {\arrow[#1]{latex}}}},
	cscalar2/.style={line width=1pt,postaction={decorate},decoration={markings,mark=at position .8 with {\arrow[#1]{latex}}}},
	photon/.style={line width =.8pt, decorate, decoration={snake, segment length=4pt, amplitude=1.8pt,  pre length=.1cm, post length=.1cm}}
}

\newcommand{\new}[1]{#1}

\makeatletter
\newlength{\apb@width}
\newcommand{\autoparbox}[2][c]{\settowidth{\apb@width}{#2}\parbox[#1]{\apb@width}{#2}}

\makeatother

\makeatletter
\def\mr@ignsp#1 {\ifx\:#1\@empty\else #1\expandafter\mr@ignsp\fi}%
\newcommand{\multiref}[1]{\begingroup
\xdef\mr@no@sparg{\expandafter\mr@ignsp#1 \: }%
\def\mr@comma{}%
\@for\mr@refs:=\mr@no@sparg\do{\mr@comma\def\mr@comma{,}\ref{\mr@refs}}%
\endgroup}
\makeatother

\renewcommand{\eqref}[1]{(\multiref{#1})}

\newcommand{\sfrac}[2]{{\textstyle\frac{#1}{#2}}}

\newcommand{\vev}[1]{\langle #1\rangle}

\newcommand{\xdot}{{\dot x}}

\newcommand{\be}{\begin{equation}}
\newcommand{\ee}{\end{equation}}
\newcommand{\ba}{\begin{align}}
\newcommand{\ea}{\end{align}}
\newcommand{\eqn}[1]{(\ref{#1})}

\newcommand{\nn}{\nonumber}
\def\dd{\delta\!\!\!{}^-\!}
\newcommand{\eps}{\varepsilon}

\usepackage{feynmp-auto}
\usepackage{feynmp} 
\newcommand{\hut}{}

\def\d{\mathrm{d}}
\def\eps{\epsilon}

\def\cO{\mathcal{O}}

\def\eps{\epsilon}

\def\bq{\mathbf q}

\def\hx{\hat{\mathbf x}}
\def\tb{\widetilde{\mathbf b}}
\def\bb{\mathbf b}
\def\bv{\hat{\mathbf e}}
\def\detTwo{\text{Det}_2}

\begin{document}

\preprint{
HU-EP-21/03-RTG
}

\title{Classical Gravitational Bremsstrahlung from a Worldline Quantum Field Theory}
\author{Gustav Uhre Jakobsen} 
\email{gustav.uhre.jakobsen@physik.hu-berlin.de}
\affiliation{%
Institut f\"ur Physik und IRIS Adlershof, Humboldt-Universit\"at zu Berlin,
Zum Gro{\ss}en Windkanal 2, 12489 Berlin, Germany
}
 \affiliation{Max Planck Institute for Gravitational Physics (Albert Einstein Institute), Am M\"uhlenberg 1, 14476 Potsdam, Germany}

\author{Gustav Mogull}
\email{gustav.mogull@aei.mpg.de} 
\affiliation{%
Institut f\"ur Physik und IRIS Adlershof, Humboldt-Universit\"at zu Berlin,
Zum Gro{\ss}en Windkanal 2, 12489 Berlin, Germany
}
 \affiliation{Max Planck Institute for Gravitational Physics (Albert Einstein Institute), Am M\"uhlenberg 1, 14476 Potsdam, Germany}

\author{Jan Plefka} 
\email{jan.plefka@hu-berlin.de}
\affiliation{%
Institut f\"ur Physik und IRIS Adlershof, Humboldt-Universit\"at zu Berlin,
Zum Gro{\ss}en Windkanal 2, 12489 Berlin, Germany
}

\author{Jan Steinhoff} 
\email{jan.steinhoff@aei.mpg.de}
 \affiliation{Max Planck Institute for Gravitational Physics (Albert Einstein Institute), Am M\"uhlenberg 1, 14476 Potsdam, Germany}


\begin{abstract}
Using the recently established formalism of a worldline quantum field theory (WQFT) description of the classical scattering of two spinless black holes, we compute the far-field time-domain waveform of the gravitational waves produced in the encounter at leading order in the post-Minkowskian (weak field, but generic velocity) expansion.
We reproduce previous results of Kovacs and Thorne in a highly economic way.
Then using the waveform we extract the leading-order total radiated angular momentum
\new{and energy (including differential results).}
Our work may enable crucial improvements of gravitational-wave
predictions in the regime of large relative velocities.
\end{abstract}

\maketitle

When two compact objects (black holes, neutron stars or stars) fly past each other their gravitational interactions not only deflect their trajectories but they also produce gravitational radiation, or gravitational Bremsstrahlung in analogy to the electromagnetic case.
The resulting waveform in the far field  at leading order in Newton's constant $G$ has been constructed (in the spinless case) in a series of papers by Kovacs, Thorne, and Crowley in the 1970s~\cite{1975ApJ...200..245T,1977ApJ...215..624C,Kovacs:1977uw,Kovacs:1978eu} ---
see Refs.~\cite{DeVittori:2012da,*Grobner:2020fnb,*Capozziello:2008mn} for recent work on slow-motion sources.
Today's gravitational wave (GW) observatories routinely detect quasi-circular \emph{inspirals and mergers} of binary black holes and neutron stars~\cite{Abbott:2016blz,*TheLIGOScientific:2017qsa,*LIGOScientific:2018mvr,*Abbott:2020niy}.
Yet Bremsstrahlung events currently appear to be out of reach as the signal is not periodic and typically less intensive~\cite{Kocsis:2006hq,*Mukherjee:2020hnm,*Zevin:2018kzq}.
Still, they represent interesting targets for GW searches, calling for accurate waveform models.

Indeed, the experimental success of GW astronomy brings up the need for high-precision theoretical predictions for the classical relativistic two-body problem~\cite{Purrer:2019jcp}.
A number of complementary classical pertubative approaches have been established over the years~\cite{Blanchet:2013haa,*Schafer:2018kuf,*Futamase:2007zz,*Pati:2000vt,*Bel:1981be,*Westpfahl:1985tsl,*Ledvinka:2008tk}.
Yet quantum-field-theory based techniques founded in a perturbative Feynman-diagrammatic expansion of the path integral in the classical limit have proven to be highly efficient.
These come in two alternative approaches.

The first approach, the effective field theory (EFT) formalism~\cite{Goldberger:2004jt,*Goldberger:2006bd,*Goldberger:2009qd}, models the compact objects as point-like massive particles coupled to the gravitational field.
It has mostly been applied to a nonrelativistic post-Newtonian (PN) scenario for \emph{bound} orbits, in which an expansion in powers of Newton's constant $G$ implies an expansion in velocities ($\frac{Gm}{c^2r} \sim \frac{v^{2}}{c^{2}}$).
Recently it has also been extended to the post-Minkowskian (PM) expansion for \emph{unbound} orbits~\cite{Kalin:2020mvi,Kalin:2020fhe} relevant for this work, an expansion in $G$ for arbitrary velocities.
In these EFT settings the graviton field $h_{\mu\nu}(x)$ is integrated out successively (from small to large length scales) in the path integral, while the worldline
trajectories of the black holes $x_i^{\mu}(\tau_{i})$ are kept as classical background sources --- see Refs.~\cite{Goldberger:2007hy,*Foffa:2013qca,*Rothstein:2014sra,*Porto:2016pyg,*Levi:2018nxp} for reviews.

The second now blossoming approach starts out from scattering amplitudes of massive scalars ---
avatars of spinless black holes ---
minimally coupled to general relativity~\cite{Neill:2013wsa,Bjerrum-Bohr:2013bxa,*Bjerrum-Bohr:2018xdl,Bern:2019nnu,*Bern:2019crd,*Cheung:2020gyp,Luna:2017dtq}, thereby putting the younger innovations in on-shell techniques for scattering amplitudes (e.g. generalized unitarity~\cite{Bern:1994zx,*Bern:1994cg,*Britto:2004nc} or the double copy \cite{Bern:2008qj,*Bern:2010ue,*Bern:2012uf,*Bern:2017ucb,*Bern:2018jmv,*Bern:2019prr}) to work.
In order to obtain the conservative gravitational potential one performs a subtle classical limit of the scattering amplitudes~\cite{Kosower:2018adc,*Maybee:2019jus,*Damour:2019lcq} in order to match to a non-relativistic EFT for scalar particles with the desired potential~\cite{Cheung:2018wkq} (see also Refs.~\cite{Neill:2013wsa,*Vaidya:2014kza,*Damour:2017zjx}), which is known to 3PM order~\cite{Bern:2019nnu,*Bern:2019crd,*Cheung:2020gyp} (complemented by certain radiation-reaction effects~\cite{Damour:2020tta, DiVecchia:2020ymx,Damour:2019lcq}).
Very recently the 4PM conservative potential was also reported \cite{Bern:2021dqo}.
The so-obtained effective potential is then used to compute observables such as the scattering angle or the (PM-resummed) periastron advance in the bound system~\cite{Kalin:2020mvi,*Kalin:2019rwq,*Kalin:2019inp}.
Further recent PM results exist for non-spinning particles~\cite{Blanchet:2018yvb,*Cristofoli:2019neg,*Cristofoli:2020uzm,*Bini:2020uiq,*Bini:2020rzn,*Loebbert:2020aos}, for spin effects~\cite{Vines:2017hyw,*Bini:2017xzy,*Bini:2018ywr,*Guevara:2017csg,*Vines:2018gqi,*Guevara:2018wpp,*Chung:2018kqs,*Guevara:2019fsj,*Chung:2019duq,*Damgaard:2019lfh,*Aoude:2020onz,*Bern:2020buy,*Guevara:2020xjx}, tidal effects~\cite{Bini:2020flp,*Cheung:2020sdj,*Haddad:2020que,*Kalin:2020lmz,*Brandhuber:2019qpg,*Huber:2019ugz,*AccettulliHuber:2020oou,*AccettulliHuber:2020oou,*Bern:2020uwk,*Cheung:2020gbf,*Aoude:2020ygw},
and radiation effects~\cite{Amati:1990xe,* DiVecchia:2019myk,* DiVecchia:2019kta,* Bern:2020gjj,* Huber:2020xny,* DiVecchia:2021ndb,*Bautista:2019tdr,*Laddha:2018rle,*Laddha:2018myi,*Sahoo:2018lxl,*Laddha:2019yaj,*Saha:2019tub,*A:2020lub,*Sahoo:2020ryf,*Sahoo:2018lxl}.

In a recent work of three of the present authors the synthesis of these two quantum-field-theory based approaches to classical relativity was provided in the form of a worldline quantum field theory (WQFT)~\cite{Mogull:2020sak}:
quantizing \emph{both} the graviton field $h_{\mu\nu}$ and the fluctuations about the bodies' worldline trajectories $z_i^{\mu}$ were shown to yield an efficient approach yielding only the relevant classical contributions. In essence the WQFT formalism provides an
efficient diagrammatic framework for solving the equations of motion of gravity-matter
systems perturbatively.

In this Letter we employ this novel formalism to compute the time-domain gravitational waveform of a Bremsstrahlung event at leading order in $G$, demonstrating its effectiveness.
To our knowledge the seminal result of Kovacs and Thorne~\cite{Kovacs:1978eu} has not been verified in its entirety to date.
As we shall see, our approach is far more efficient than the one employed back then, paving the way for calculations of higher orders.
We stress that we are able to determine the \emph{far-field waveforms} which are of direct relevance for GW observatories.
As a check on these waveforms we furthermore
reproduce Damour's recent result for the \emph{total radiated
angular momentum} \cite{Damour:2020tta} at 2PM order.
Our results also complement the recent result of the
\emph{total radiated momentum} at leading order in
$G$ (3PM) established with amplitude techniques~\cite{Herrmann:2021lqe}.
We comment on how to achieve this result from our methods.

 \begin{figure}[t!]
  \includegraphics[width=8cm]{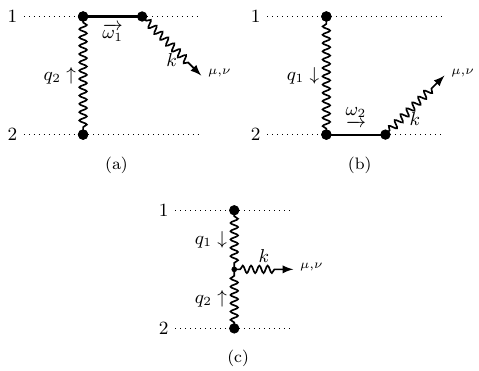}
    \caption{The three diagrams contributing to the Bremsstrahlung at 2PM order,
    where $\omega_i=k\cdot v_i$ by energy conservation at the worldline vertices.
    All three diagrams have the integral measure in Eq.~\eqn{eq:measure};
    in the rest frame of black hole 1 diagram (a) does not contribute as soon as the outgoing graviton is contracted with a purely spatial polarization tensor.}
  \label{fig:1}
   \end{figure}

\sec{Worldline Quantum Field Theory}
The classical gravitational 
scattering of two massive objects $m_{i}$ moving on trajectories
$x^\mu_{i}(\tau_{i})=b^\mu_{i}+v^\mu_{i}\tau_i +z^\mu_{i}(\tau_{i})$
is described by the worldline quantum field theory (WQFT) with partition function 
\cite{Mogull:2020sak}
\begin{align}\label{ZWQFTdef}
\mathcal{Z}_{\text{WQFT}}
&:= \text{const} \times
\int \!\!D[h_{\mu\nu}]
\int \prod_{i=1}^{2} D [z_{i}] \,  e^{i (S_{\rm EH}+S_{\rm gf})}\\ &\quad
\exp\Bigl[ -i\sum_{i=1}^{2}\int_{-\infty}^{\infty}\!\d\tau_{i}\, \frac{m_{i}}{2}
[\eta_{\mu\nu}+\kappa h_{\mu\nu}(x)]\xdot_{i}^{\mu}\,\xdot_{i}^{\nu}\Bigr ],\nn
\end{align}
where $S_{\rm EH}+S_{\rm gf}$ is the gauge-fixed Einstein-Hilbert action
\begin{align}\label{eq:einsteinHilbert}
S_{\rm EH}+S_{\rm gf}&=\int\!\d^4x\left ( -\sfrac{2}{\kappa^{2}}\sqrt{-g}R\,
+(\partial_\nu h^{\mu\nu}-\sfrac12\partial^\mu{h^\nu}_\nu)^2 \right )\, ,
\end{align}
with $\kappa^{2}= 32\pi G$ the gravitational coupling;
we have suppressed the ghost contributions in Eq.~\eqn{ZWQFTdef} as they are irrelevant in the classical setting.
We work in mostly minus signature, $\eta_{\mu\nu}={\rm diag(1,-1,-1,-1)}$\new{, and set $c=\hbar=1$}.

Correlation functions in the WQFT $\langle \cO(h,\{x_i\}) \rangle_{\text{WQFT}}$ result from an insertion of the operator $\cO$ in the path integral and dividing by $\mathcal{Z}_{\text{WQFT}}$. Moving to momentum space for the graviton $h_{\mu\nu}(k)$ and energy space for the fluctuations $z^{\mu}(\omega)$ we have the \emph{retarded} propagators
\begin{subequations}\label{eq:Propagators}
\begin{align}
  \raisebox{-0.3cm}{\includegraphics[width=1.6cm]{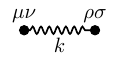}}
  &=i\frac{P_{\mu\nu;\rho\sigma}}{(k^{0}+i\epsilon)^{2}-\mathbf{k}^2}\,, \,\\ 
\raisebox{-0.3cm}{\includegraphics[width=1.6cm]{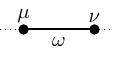}}&=-i\frac{\eta^{\mu\nu}}{m\,(\omega+i\eps)^2}\,,
\end{align}
\end{subequations}
with $P_{\mu\nu;\rho\sigma}:=\eta_{\mu(\rho}\eta_{\sigma)\nu}-
\sfrac12\eta_{\mu\nu}\eta_{\rho\sigma}$.
The relevant vertices for the emission of a graviton off the worldline read
\begin{align}\label{eq:vertexH}
 \raisebox{-1.0cm}{\includegraphics[width=1.2cm]{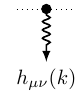}}
 =
  -i\frac{m\kappa}{2}e^{ik\cdot b}\dd(k\cdot v)v^\mu v^\nu\,,
  \end{align}
  with $k$ outgoing, $\dd(\omega):=(2\pi)\delta(\omega)$ and
  \begin{align}\label{eq:vertexHZ}
 \raisebox{-1.0cm}{\includegraphics[width=2.0cm]{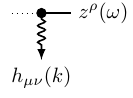}}&=
  \frac{m\kappa}{2}e^{ik\cdot b}\dd(k\cdot v+\omega)
  \qquad\qquad
\end{align}
\vspace{-4em}
\begin{align*}
  \qquad\qquad\qquad\qquad\qquad\qquad
  \times\left(2\omega v^{(\mu}\delta^{\nu)}_\rho+v^\mu v^\nu k_\rho\right)\,.
\end{align*}
The energy $\omega$ is also taken as outgoing.
One also has the standard bulk graviton vertices,
of which we shall need only the three-graviton vertex ---
see e.g.~Ref.~\cite{Sannan:1986tz}.

To determine the Bremsstrahlung of two traversing black holes 
we compute the expectation value $k^2\vev{h^{\mu\nu}(k)}_{\rm WQFT}$.
At leading (2PM) order there are three diagrams contributing, cp.~Fig.~\ref{fig:1}.
We integrate over the momenta or energies
of internal gravitons or fluctuations respectively;
lack of three-momentum conservation at the worldline vertices
leaves unresolved integrals for the tree-level diagrams.
\begin{widetext}
Diagram (a) of Fig.~\ref{fig:1} then takes the form\footnote{In
	principle we should also contract with $P_{\mu\nu;\rho\sigma}$ for
	an outgoing graviton line; however as the polarization tensors
	${e^{\mu\nu}_{+,\times}}$ are traceless we find it unnecessary.
}
\be
k^2\vev{h_{\mu\nu}(k)}_{\rm WQFT}\Bigr|_{(a)} \!= -\frac{m_1m_2\kappa^{3}}{8}\!
\int_{q_1,q_2}{\mu}_{1,2}(k)
\frac{(2\omega_1 {v}_1^{(\mu}\delta^{\nu)}_\rho-{v}_1^\mu {v}_1^\nu k_\rho)
(2\omega_1 {v}_1^{(\sigma}\eta^{\lambda)\rho}-{v}_1^\sigma {v}_1^\lambda q_2^\rho)}
{(\omega_1+i\epsilon)^2}
\frac{P_{\sigma\lambda;\alpha\beta}}{[(q_2^{0}+i\epsilon)^{2}-\mathbf{q_{2}}^{2}]}{v}_2^\alpha {v}_2^\beta,
\ee
where $\omega_1=k\cdot v_1$,
$\int_{q_i}:=\int\!\!\sfrac{{\rm d}^4q_i}{(2\pi)^4}$ and the integral measure is
\begin{equation}\label{intmeasureGold}
\hut{\mu}_{1,2}(k)=
e^{i(q_1\cdot\hut{b}_1+q_2\cdot\hut{b}_2)}
\dd(q_1\cdot\hut{v}_1)\dd(q_2\cdot\hut{v}_2)
\dd(k-q_1-q_2)\,,
\end{equation}
with $\dd(k):=(2\pi)^4\delta^{(4)}(k)$. The diagram (b) is naturally obtained by swapping 
$1\leftrightarrow 2$. Diagram (c) includes the three-graviton vertex
$V_3^{(\mu\nu)(\rho\sigma)(\lambda\tau)}(k,-q_1,-q_2)$:
\be
k^2\vev{h_{\mu\nu}(k)}_{\rm WQFT}\Bigr|_{(c)}=-\frac{m_1m_2\kappa^{3}}{8}
\int_{q_1,q_2}{\mu}_{1,2}(k)
V_3^{(\mu\nu)(\rho\sigma)(\lambda\tau)}
\frac{P_{\rho\sigma;\alpha\beta}}{[(q_1^{0}+i\epsilon)^{2}-\mathbf{q_{1}}^{2}]}
\frac{P_{\lambda\tau;\gamma\delta}}{[(q_2^{0}+i\epsilon)^{2}-\mathbf{q_{2}}^{2}]}
{v}^\alpha_1{v}^\beta_1{v}^\gamma_2{v}^\delta_2\,.
\ee
\end{widetext}
These integrands were already given in Ref.~\cite{Mogull:2020sak}.
The sum of the three integrands also agrees with a previous amplitudes-based
result \cite{Luna:2017dtq} \new{(see also Ref.~\cite{Goldberger:2016iau} for the analogue in dilaton gravity)} and is gauge-invariant.

The waveform in spacetime in the \emph{wave zone} is obtained from
$\vev{h^{\mu\nu}(k)}_{\rm WQFT}$ as follows:
we may identify
\be
k^2\vev{h_{\mu\nu}(k)}_{\rm WQFT} = \frac\kappa2\,S_{\mu\nu}(k)\,,
\ee
where $S_{\mu\nu}= \tau_{\mu\nu}-\sfrac{1}{2}\eta_{\mu\nu}\tau^{\lambda}{}_{\lambda}$ and
$\tau_{\mu\nu}$ is the combined energy-momentum pseudo-tensor of matter and the gravitational
field. Consider $S_{\mu\nu}(k)$ for a fixed GW frequency $k^{0}=\Omega$.
In the wave zone ($r\gg \{|b_{i}|,\Omega^{-1},\Omega|b_{i}|^{2}\}$)
the metric perturbation $h_{\mu\nu}(\mathbf{x},t)$ takes the form of a plane wave
(see e.g.~Chapter 10.4 of Weinberg~\cite{Weinberg:1972kfs}):
\be
\kappa h_{\mu\nu}(\mathbf{x},t)=
\frac{4G}{r}\,S_{\mu\nu}(\Omega,\mathbf{k}=\Omega\,\mathbf{\hat{x}})\,
e^{-ik_{\mu}x^{\mu}} +c.c\,,
\ee
with the wave vector $k^{\mu}=\Omega(1,\mathbf{\hat x})$;
$\mathbf{\hat x}=\mathbf{x}/r$ is the unit vector pointing in the direction of the observation point (hence $k^2=0$). 

The total gauge-invariant frequency-domain waveform can be read off as $4G \, S_{ij}^\text{TT}(\Omega,\mathbf{k}=\Omega\,\mathbf{\hat{x}})$,
where TT denotes the transverse-traceless projection.
The corresponding time-domain waveform $f_{ij}(u,\theta,\phi)$
is essentially its Fourier transform in $\Omega$:
\be
\label{eq:startingpoint}
\kappa h_{ij}^{\rm TT}=\frac{f_{ij}}{r} = \frac{4G}{r}
\int_\Omega e^{-i k\cdot x}
S_{ij}^\text{TT}(k) \Bigr |_{k^{\mu}=\Omega\,(1,\hat{\bf x})}\, ,
\ee
where $\int_\Omega:=\int_{-\infty}^\infty\sfrac{{\rm d}\Omega}{2\pi}$.
Note that $k\cdot x = \Omega (t-r)$ yields the retarded time $u=t-r$.
Our task now is to perform the integrals;
in a PM expansion $f_{ij}=\sum_nG^nf_{ij}^{(n)}$,
and we seek the 2PM component $f_{ij}^{(2)}$.
By focusing on the time-domain instead of the frequency-domain waveform
we considerably simplify the integration step ---
as we shall see, the integration over frequency $\Omega$ of the outgoing
radiation coincides neatly with energy conservation along each worldline.

\sec{Kinematics}
We describe the waveform in a Cartesian coordinate system $(t,x,y,z)$ where black hole 1 is initially at rest $v_{1}^{\mu}=(1,0,0,0)$ and located at the spatial origin, i.e.~we set $b_{1}^{\mu}=0$.
The orbit of black hole 2 we put in the $x$--$y$ plane with initial velocity 
$v_{2}^{\mu}=(\gamma,\gamma\, v,0,0)$ in the $x$-direction;
the impact parameter  $b_{2}^{\mu}=(0,0,b,0)=:b^{\mu}$ points in the
$y$-direction.
Introducing the polar angles $\theta$ and $\phi$ 
we may write the unit (spatial) vector $\hat{x}^{\mu}$ pointing from black hole 1 to
our observation point as
\be
\hat{x}^\mu
  =
  \hat{e}_1^\mu \cos\theta 
  +
  \sin\theta
  \big(
  \hat{e}_2^\mu\cos\phi 
  +
  \hat{e}_3^\mu\sin\phi 
  \big)\, ,
\ee
where $\hat{e}_i^\mu=(0,\bv_i)$ are spatial unit vectors.
Also, we put $\rho^{\mu} = v_1^{\mu} + \hat{x}^{\mu}$.

The two additional unit spatial vectors orthogonal to $\hat{x}^\mu$ are
\be
  \hat{\theta}^{\mu}
  =
  \partial_\theta \hat{x}^{\mu}=(0,\hat{\bm\theta})
  \ , \quad
  \hat{\phi}^{\mu}
  =
  \sfrac{1}{\sin\theta} \partial_\phi \hat{x}^{\mu}=(0,\hat{\bm\phi})\, .
\ee
Together with $\hat{x}^\mu$ they
form a right-handed spatial coordinate system.
GWs travel in the direction of $\hat{x}^\mu$ and we use 
$\hat{\theta}^\mu$ and $\hat{\phi}^\mu$ to define
our polarization tensors in a linear basis:
\begin{align}
  e_+^{\mu\nu}
  =
  \hat{\theta}^\mu \hat{\theta}^\nu - \hat{\phi}^\mu \hat{\phi}^\nu\,,
  \quad
  e_\times^{\mu\nu}
  =
  \hat{\theta}^\mu \hat{\phi}^\nu
  +
  \hat{\phi}^\mu \hat{\theta}^\nu\, .
\end{align}
The waveform $f_{ij}(u,\theta,\phi)$ is thus decomposed as
\begin{equation}\label{eq:polBasis}
f_{ij}=f_+ (e_+)_{ij}+f_\times(e_\times)_{ij} \,
\end{equation}
with $f_{+,\times}=\frac12(e_{+,\times})_{ij}f_{ij}$.

The polarization tensors have zero time components,
which conveniently implies the vanishing of diagram (a)
in Fig.~\ref{fig:1} once contracted with them.
This observation follows directly from the expression for vertex \eqref{eq:vertexHZ}:
in the case of diagram (a) the instance of this vertex that
contracts with the outgoing graviton line carries
an overall factor of $v_1^\mu=(1,0,0,0)$,
which is orthogonal to the spatial polarization tensors above.

\sec{Integration}
The two non-zero diagrams in Fig.~\ref{fig:1}
share the integration measure $\mu_{1,2}(k)$ \eqref{intmeasureGold}.
Including also the integration with respect to $\Omega$
in Eq.~\eqref{eq:startingpoint} the full measure becomes
\begin{align}\label{eq:measure}
\int_{\Omega,q_1,q_2}\mu_{1,2}(k)e^{-ik\cdot x}=
\frac1{\rho\cdot v_2}\int_\bq
e^{i\bq\cdot\widetilde{\mathbf b}}\,,
\end{align}
where we recall that $k^\mu=\Omega\rho^\mu$;
using the delta function constraints in $\mu_{1,2}(k)$ we can now identify
\begin{align}
q_2=k-q_1\,, &&
q_1=(0,\bq)\,, &&
\Omega=-\frac{v\gamma}{\rho\cdot v_2}\bq\cdot\bv_1\,.
\end{align}
We are left with a three-dimensional Euclidean integral involving 
the shifted $\tau$-dependent impact parameter:
\begin{align}
\tb(\tau)&={\mathbf b}+\tau\,\bv_1\,,\ &&
\tau
= \frac{v\gamma}{\rho\cdot v_2}
(u+\mathbf{b}\cdot\mathbf{\hat x})\,,
\end{align}
noting that $\rho\cdot v_{2}= \gamma(1- v \cos\theta)$.
The polarizations of the waveform from Eq.~\eqref{eq:startingpoint} now take the schematic form (also using Eq.~\eqref{eq:polBasis})
\begin{align}\label{eq:schematic}
  &\frac{f^{(2)}_{+,\times}}{m_1 m_2}\\
  &=
  4\pi \int_\bq
  e^{i\bq\cdot\tb}\left(
  \frac{{\mathcal N}^i_{+,\times}\bq^i}{\bq^2(\bq\cdot\bv_1-i\eps)}+
  \frac{{\mathcal M}^{ij}_{+,\times}\bq^i\bq^j}{\bq^2(\bq^2+\bq\cdot L\cdot\bq)}
  \right)\,,\nn
\end{align}
with the two terms corresponding to the non-zero diagrams
(b) and (c) in Fig.~\ref{fig:1} respectively.
The rank-2 matrix $L$ introduced here is
\begin{equation}\label{eq:Lmatrix}
L^{ij}=2\frac{v\gamma}{\rho\cdot v_2}\bv_1^{(i}\hx^{j)}\,.
\end{equation}
Finally the vector and matrix insertions are explicitly given as the real and imaginary parts of\footnote{To
  compactify these results we have used the generalized gauge invariance
  ${\mathcal N}_{+,\times}\to{\mathcal N}_{+,\times}+X\hat{\mathbf e}_1$,
  ${\mathcal M}_{+,\times}\to{\mathcal M}_{+,\times}-X(\mathbb{1}+L)$
  for an arbitrary function $X$ of external kinematics.
  We have also dropped a term from ${\cal N}_{\pm}$ in the $\bv_3$
  direction which does not contribute to the final integrated result~\eqref{eq:result}.
}
\begin{subequations}
  \begin{align}
    {\mathcal N}^i&=
    2\frac{ \gamma^2\sin^2{\theta}}{\rho\cdot v_2}
    \Big(
    \frac{\gamma(1-3v^2)}{\rho\cdot v_2}
    + (1+v^2)
    \Big)
    \bv_1^i
    \\
    &
    +2\frac{\gamma(1+v^2)\sin{\theta}}{\rho\cdot v_2}
    \Big(
    \frac{(\rho\cdot v_2)^2-1}{v(\rho\cdot v_2)}
    \cos{\phi}
    + 2i\gamma
    \sin{\phi}
    \Big)
    \hat{\bm e}_2^i,
    \nn
    \\
    {\mathcal M}^{ij}&=
    8\frac{\gamma^4v^4\sin^2\theta}{(\rho\cdot v_2)^3}\bv_1^i\bv_1^j+
    16\frac{\gamma^3v^2\sin\theta}{(\rho\cdot v_2)^2}\bv_1^{(i}
    (\hat{{\bm \theta}}+ i\hat{{\bm \phi}})^{j)}
    \nn
    \\
    &\qquad
    +
    4\frac{\gamma^2(1+v^2)}{\rho\cdot v_2}
    (\hat{{\bm \theta}}+ i\hat{{\bm \phi}})^i
    (\hat{{\bm \theta}}+ i\hat{{\bm \phi}})^j\,,
\end{align}
\end{subequations}
where ${\mathcal N}^i= {\mathcal N}^i_+ + i{\mathcal N}^i_\times$
and ${\mathcal M}^{ij}= {\mathcal M}^{ij}_+ + i{\mathcal M}^{ij}_\times$.
The insertions ${\mathcal N}^i$ and ${\mathcal M}^{ij}$
correspond to a helicity basis in which they have a particular simple expression.
We integrate the two diagrams separately.

Integration of the first diagram is achieved using the simple result
(true regardless of the vector $\tb$)
\begin{align}\label{eq:easyInt}
&\int_\bq e^{i\bq\cdot\tb}
\frac{\bq^i}{\bq^2(\bq\cdot\bv_1-i\eps)}\\
&\quad=
\frac1{4\pi}\left(
	\frac{\bv_1^i}{|\tb|}-
	\frac{\tb^i-(\tb\cdot\bv_1)\bv_1^i}
	{\tb^2-(\tb\cdot\bv_1)^2}
	\left(1+\frac{\tb\cdot\bv_1}{|\tb|}\right)
\right)\,,\nn
\end{align}
which we prove in the Appendix.
The other integral required corresponding to diagram (c) is
somewhat more involved.
The denominator of this integral is composed of an isotropic propagator together with an anisotropic one.
The physical interpretation is a convolution between the potentials of the two black holes, where the potential of black hole 2 is boosted and leads to the anisotropic propagator.
One compact representation is
\begin{align}\label{eq:hardInt}
  &\int_\bq e^{i\bq\cdot\tb}
  \frac{\bq^i\bq^j
  }{
    \bq^2(\bq^2+\bq\cdot L \cdot\bq)}
  \\
  &=
  \frac{1}{2\pi \Delta(G)}
    \left[
    \frac{(G_0+\alpha G_1)A^{ij}-(G_1+\alpha G_2)B^{ij}}{\sqrt{G(\alpha)}}
    \right]_{\alpha=0}^{\alpha=1}\,,\nn
\end{align}
where we have introduced the quadratic polynomial
\begin{align}\label{eq:gFunction}
  &G(\alpha)=G_0+2\alpha G_1+\alpha^2G_2
  \,,
  \\
  &
  G_0=\tb^2
  \,,
  G_1=
  \tb^i\tb^j\frac{\delta^{ij} L^{kk}- L^{ij}}{2}
  \,,
  G_2=
  (\tb\cdot\hat{\bm \phi})^2
  \detTwo{L}
  \,,
  \nn
\end{align}
and $\Delta(G)=4(G_1^2-G_0G_2)$ is the polynomial discriminant.
We have also introduced the two matrices
\begin{subequations}
  \begin{align}
    A^{ij}&=
    \detTwo(L)
    \Big(
    -2(\tb\cdot\hat{\bm\phi})(L^{-1}\cdot\tb)^{(i}\hat{\bm\phi}^{j)}
    \\
    &\qquad+(\tb\cdot\hat{\bm\phi})^2(L^{-1})^{ij}
    +(\tb\cdot L^{-1} \cdot\tb)\hat{\bm\phi}^i\hat{\bm\phi}^j
    \Big)
    \,,\nn\\
    B^{ij}&=
    \tb^2\delta^{ij}-\tb^i\tb^j\,.
  \end{align}
\end{subequations}
This integral is also discussed in the Appendix where explicit forms
of $L^{-1}$ and $\detTwo(L)$ are given;
again, both of the integrals \eqref{eq:easyInt} and \eqref{eq:hardInt}
are solved for arbitrary $\tb^i$ and $L^{ij}$
(with the assumption that $L^{ij}$ is rank 2).

\sec{Leading-Order Waveform}
By combining Eq.~\eqref{eq:schematic} with the insertions $\mathcal{N}_{+,\times}^i$ and $\mathcal{M}_{+,\times}^{ij}$ and the integrals above we get the full 2PM waveform:
\begin{align}\label{eq:result}
\begin{aligned}
  &\frac{f^{(2)}_{+,\times}}{m_1 m_2}
  =
  \frac{
    \bv_1^i \mathcal{N}^i_{+,\times}
  }{
    \sqrt{\bb^2+\tau^2}}
  -
  \frac{
    \bb^i \mathcal{N}^i_{+,\times}
  }{
    \bb^2}
  \left(1+\frac{\tau}{\sqrt{\bb^2+\tau^2}}\right)
  \\
  &
  \!+
  \frac{2\mathcal{M}^{ij}_{+,\times}}{\Delta(G)}
  \bigg[
    \frac{
    (G_0+\alpha G_1) A^{ij}
    -
    (G_1+\alpha G_2) B^{ij}
    }{
      \sqrt{G(\alpha)}}
    \bigg]_{\alpha=0}^{\alpha=1}\!\!.
\end{aligned}
\end{align}
This is a rather compact representation of the gravitational Bremsstrahlung waveform, which we have confirmed agrees with
the (rather lengthy) result of Kovacs and Thorne~\cite{Kovacs:1978eu}.
The two values of $\alpha$ in the second line correspond to contributions from the two black holes.
Note that there is also a leading (and non-radiating)
1PM contribution to the waveform which is independent of the retarded time $u=t-r$:
\begin{align}\label{eq:leading}
  f_+^{(1)}=\frac{2m_2\gamma v^2\sin^2\theta}{1-v\cos\theta}\,, &&
  f_\times^{(1)}=0\,.
\end{align}
Diagrammatically this consists only of the vertex \eqref{eq:vertexH}
with emission from worldline 2;
the contribution from worldline 1 again vanishes
in our frame due to $(v_{1}\cdot e_{\pm}\cdot v_{1})=0$.

\begin{figure}[t!]
  \includegraphics[width=4cm]{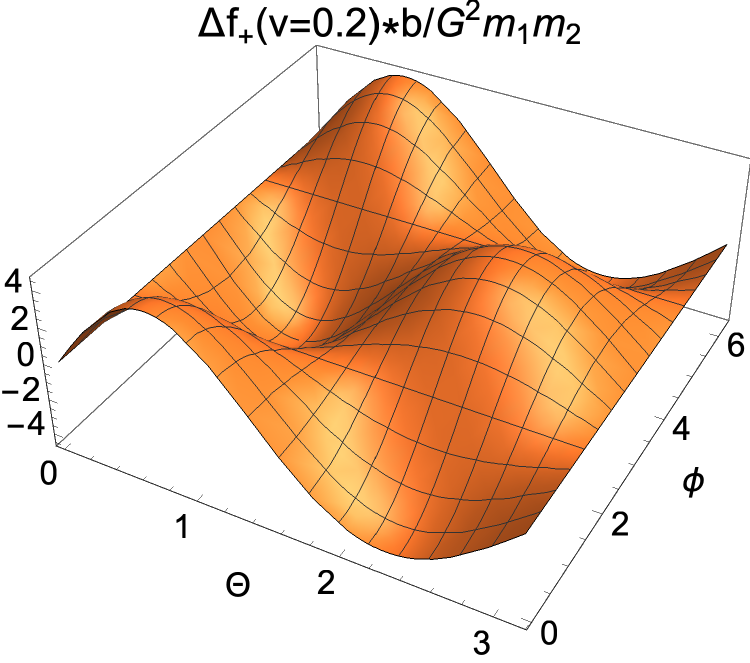}
  \includegraphics[width=4cm]{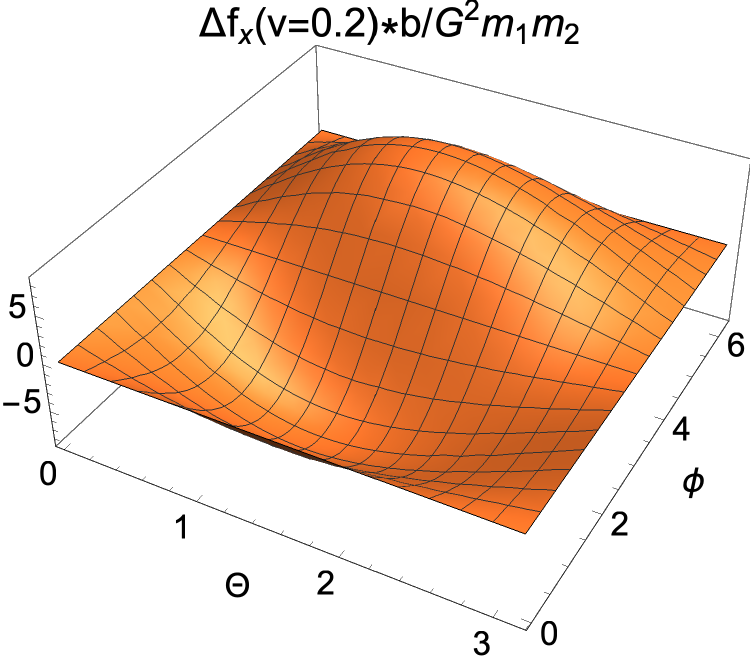}
   \caption{Plots of the wave memories $\Delta f_{+,\times}$ for $v=0.2$. For a visualisation of the complete waveforms as they evolve with retarded time $u$ see 
  \href{https://box.hu-berlin.de/f/73b510f4dae1408ba612/}{$f_{+}(u,\theta,\phi)|_{v=0.2}$} and 
  \href{https://box.hu-berlin.de/f/6c1ff4812f7448ca913d/}{$f_{\times}(u,\theta,\phi)|_{v=0.2}$}.}
  \label{fig:2}
\end{figure}

To illustrate this result in Fig.~\ref{fig:2} we present the
\emph{gravitational wave memories}
$\Delta{f}_{+,\times} := [ {f}_{+,\times} ]^{u=+\infty}_{u=-\infty}$.
The beauty of our result \eqn{eq:result} lies in the fact that the memories
only receive contributions from the second term, and read
\be
\Delta f_{+,\times}=
-2G^2m_1m_2
  \frac{
    \bb^i \mathcal{N}^i_{+,\times}
  }{
    \bb^2}+\cO(G^3)\, .
\ee
Diagrammatically they exclusively emerge from diagram (b) of Fig.~\ref{fig:1}.
So they are manifestly insensitive to gravitational self-interactions ---
this was also pointed out in Ref.~\cite{Damour:2020tta}.

\sec{\new{Radiated Energy and Angular Momentum}}
One may now use our result for the waveform \eqn{eq:result} to compute the total radiated momentum and angular momentum.
Expressions for these quantities in terms of the asymptotic waveform are given in
Refs.~\cite{Bonga:2018gzr,Damour:2020tta}:
\begin{align}
P^{\mu}_{\text{rad}}&=
\frac1{32\pi G}\int\!\d u \d\sigma [\dot{f}_{ij}]^{2}\rho^{\mu}\label{eq:momdef}\,,\\
J^{\text{rad}}_{ij}&=\frac1{8\pi G}\int\!\d u \d\sigma \left (f_{k[i}\dot{f}_{j]k} -\frac{1}{2} x_{[i}\partial_{j]}f_{kl}\dot{f}_{kl}\right )\, ,
\end{align}
where $\dot{f}_{ij}:=\partial_{u}f_{ij}$ and
$\d\sigma=\sin\theta\d\theta\d\phi$ is the unit sphere measure.

\new{
We first concentrate on $J^{\text{rad}}_{ij}$
as it contributes at leading order $\mathcal{O}(G^2)$ and 
was recently obtained in the center-of-mass frame \cite{Damour:2020tta}.}
The static nature of $f^{(1)}_{ij}$ \eqref{eq:leading} allows
one to trivially perform the $u$-integration and express the radiated
angular momentum in terms of the wave memories $\Delta{f}_{+,\times}$.
Inserting the basis of polarization tensors \eqref{eq:polBasis} 
(and using $f^{(1)}_\times=0$) gives
\begin{align} \label{eq:sphi}
  J^{\text{rad}}_{xy}&=\frac1{8\pi}\int\!\d\sigma\Bigl [ 
  \frac{\sin\phi}{\sin\theta} f_{+}^{(1)}\Delta f_{\times}-
  \frac12\cos \phi \,\partial_{\theta}f_{+}^{(1)}\, \Delta f_{+} 
  \Bigr ]\nn\\
  &\qquad+\cO(G^3)\,.
\end{align}
The spherical integral is elementary and yields
\begin{subequations}
\begin{align}
\frac{J^{\text{rad}}_{xy}}{J^{\text{init}}_{xy}}&=
\frac{4G^2m_1m_2}{b^2}\frac{(2\gamma^2-1)}{\sqrt{\gamma^2-1}}{\cal I}(v)+\cO(G^3)\,,\\
{\cal I}(v)&=-\frac83+\frac1{v^2}+\frac{(3v^2-1)}{v^3}{\rm arctanh}(v)\,,
\end{align}
\end{subequations}
where we have normalized our result with respect to
the initial angular momentum in the rest frame of black hole 1:
$J^{\text{init}}_{xy}=m_2|{\mathbf v}_2||\bb|=m_2\gamma vb$.
\new{We find perfect agreement with Ref.~\cite{Damour:2020tta}.\footnote{As the two frames are related by a boost in the $x$ direction this implies that $J_{0y}^{\text{rad}}=0$ in both frames.}

Similarly, $P^{\mu}_{\text{rad}}$ of Eq.~\eqn{eq:momdef} should reproduce
the recent result of Ref.~\cite{Herrmann:2021lqe} contributing at $\mathcal{O}(G^3)$. 
So far we have only been able to perform the integral in the PN expansion recovering the 
result of Ref.~\cite{Herrmann:2021lqe} to order $v^{6}$. Yet it is straightforward to obtain differential
quantities derived from the integrand of Eq.~\eqn{eq:momdef}. 
The differential power spectrum (total energy radiated per unit frequency)
as well as the total energy radiated per unit solid angle 
are collected in the supplementary material
to this letter. These results go well beyond Kovacs and Thorne \cite{Kovacs:1978eu} and may be expanded to any desired order in $v$.}

\new{\sec{Conclusions}}
Searching for GWs from scattering events over the full range of impact velocities requires precision predictions in the PM approximation.
While the potential and radiation of bound systems was calculated to high PN order~\cite{Damour:2014jta,*Damour:2016abl,*Bernard:2016wrg,*Foffa:2016rgu,*Damour:2017ced,*Foffa:2019rdf,*Foffa:2019yfl,*Blumlein:2020pog,*Porto:2017dgs,*Marchand:2017pir,*Galley:2015kus,*Foffa:2019hrb,*Blumlein:2019zku,*Bini:2019nra,*Blumlein:2020pyo,*Blumlein:2020znm,*Bini:2020nsb,*Bini:2020wpo,*Bini:2020hmy,*Blumlein:2021txj,*Blanchet:2001aw,*Blanchet:2004ek,*Blanchet:2008je} (see Refs.~\cite{Levi:2020kvb,*Antonelli:2020aeb,*Levi:2016ofk,*Levi:2020uwu,*Levi:2019kgk,*Levi:2020lfn,*Leibovich:2019cxo,*Mishra:2016whh,*Buonanno:2012rv,*Porto:2010zg,*Porto:2012as,*Maia:2017gxn,*Maia:2017yok} for spinning bodies), a resummation of PN results in the strong-field and fast-motion
regimes is essential for building accurate waveform models~\cite{Purrer:2019jcp}.
The PM resummation is one promising recent attempt~\cite{Damour:2016gwp,Damour:2017zjx,Antonelli:2019ytb}.

Our results provide a stepping stone for higher-order calculations,
where a repertoire of advanced integration techniques can be put to use~\cite{Bern:2019crd,Parra-Martinez:2020dzs,Bern:2021dqo,Herrmann:2021lqe}.
\new{In fact the 3PM integrand has essentially been presented in Ref.~\cite{Mogull:2020sak}. The present challenge lies in the multi-scale integrals, which despite their tree-level structure are of higher loop three-momentum type as the worldline only  preserves energy. }
Generalizations to spin and finite-size effects \new{are} possible
\new{and lead to the same families of integrations at 2PM}. Also the extensions to \emph{bound} systems using mappings between bound and unbound orbits  \cite{Kalin:2019rwq,Kalin:2019inp,Bini:2012ji} 
would be of great utility.

\sec{Acknowledgments}
We would like to thank A.~Buonanno and J.~Vines for helpful discussions.
We are also grateful for use of G.~K\"alin's C\texttt{++} graph library.
GUJ's and GM's research is funded by the Deutsche Forschungsgemeinschaft (DFG, German Research Foundation) Projektnummer 417533893/GRK2575 ``Rethinking Quantum Field Theory''.

\section*{Supplemental Material}

\sec{Integrals}\label{sec:integrals}
We begin with the simpler integral in Eq.~\eqref{eq:easyInt},
corresponding to diagram (b) in Fig.~\ref{fig:1}.
Working in Cartesian components with $\bb=(b_1,b_2,b_3)$
(in the main text we replace $\bb\to\tb$)
and $\bq=(q_1,q_2,q_3)$ it is sufficient to show
\begin{align}\label{eq:easyTemplate}
\int_\bq e^{i\bq\cdot\bb}
\frac{q_2}{\bq^2(q_1-i\eps)}
=-\frac{b_2}{4\pi(\bb^2-b_1^2)}
\left(1+\frac{b_1}{|\bb|}\right).
\end{align}
The corresponding result with numerator $q_3$ is related by symmetry,
and the one with $q_1$ is trivially given by
\begin{equation}\label{eq:gauss}
\int_\bq\frac{e^{i\bq\cdot\bb}}{\bq^2}=\frac1{4\pi|\bb|}\,.
\end{equation}
We make convenient use of the fact that
\begin{equation}
\int_\omega e^{i\omega\tau}\frac{f(\omega)}{\omega-i\eps}=
i\int_{-\infty}^\tau\!{\rm d}\tau'\int_\omega e^{i\omega\tau'}f(\omega)\,,
\end{equation}
the $i\eps$ prescription implying that our integrals
are boundary fixed at $\tau\to-\infty$.
This is to be expected given our use of retarded propagators.
The original integral can therefore be re-written as
\begin{align}
\int_\bq e^{i\bq\cdot\bb}\frac{q_2}{\bq^2(q_1-i\eps)}=
\frac{\partial}{\partial b_2}\int_{-\infty}^{b_1}\!{\rm d}b'_1
\int_\bq e^{i\bq\cdot\bb'}\frac1{\bq^2}\,,
\end{align}
where $\bb'=(b'_1,b_2,b_3)$.
From here using Eq.~\eqref{eq:gauss}
it is a simple exercise to reproduce Eq.~\eqref{eq:easyTemplate}.

Next, let us consider the anisotropic integral \eqref{eq:hardInt}
corresponding to the gravitational self-interaction of the two black hole potentials.
The numerator can be obtained by differentiation with respect to the impact parameter;
also introducing a Feynman parameter $\alpha$ the integral is re-expressed as
\begin{align}
\begin{aligned}
  {\cal I}^{ij}&=\int_\bq e^{i\bq\cdot\bb}
  \frac{\bq^i\bq^j
  }{
    \bq^2(\bq^2+\bq\cdot L \cdot\bq)}\\
  &=
  -\frac{\partial}{\partial\bb^i}
  \frac{\partial}{\partial\bb^j}
  \int_0^1{\rm d}\alpha \int_\bq
  \frac{e^{i\bq\cdot\bb}}{
  (\bq^2+\alpha\bq\cdot L\cdot\bq)^2}\,.
\end{aligned}
\end{align}
The $\bq$-integration and $\bb$ derivatives are straightforward;
we are left with the Feynman parameter integral
\begin{align}\label{eq:FeynmanPar}
  {\cal I}^{ij}=
  \int_0^1\!{\rm d}\alpha
  \frac{
    \det(M) \bb^k \bb^l
    (M^{-1})^{k[l} (M^{-1})^{i]j}
  }{
    4\pi G(\alpha)^{3/2}}\,,
\end{align}
where the quadratic function $G(\alpha)=\det(M)\bb \cdot M^{-1} \cdot\bb$
was given in Eq.~\eqref{eq:gFunction} and the $\alpha$-dependent matrix $M^{ij}$
is defined as simply
\begin{align}
  M^{ij}=\delta^{ij}+\alpha L^{ij}\,.
\end{align}
The Feynman parameter integral is solved by recognizing that the numerator in Eq.~\eqref{eq:FeynmanPar} is a linear function in $\alpha$: $B^{ij}+\alpha A^{ij}$,
where all dependence on $\alpha$ is explicit.
When the numerator in Eq.~\eqref{eq:FeynmanPar} is rewritten
in this way the Feynman parameter integral can
easily be solved to give Eq.~\eqref{eq:hardInt}.

In the present case where $L^{ij}$ is given by Eq.~\eqref{eq:Lmatrix} we find that
\begin{subequations}
\begin{align}
  \detTwo(L) &= -
  \Big(
  \frac{\gamma v \sin\theta}{\rho \cdot v_2}
  \Big)^2\,,
  \\
  (L^{-1})^{ij}
  &=
  \frac{\rho\cdot v_2}{2 \gamma v}
  \sum_{\pm}
  \pm
  \frac{
    ( \hat {\bm x} \pm \bv_1 )^i
    ( \hat {\bm x} \pm \bv_1 )^j
  }{
    (1 \pm \cos{\theta})^2
  }\,,\label{eq:L-1}
\end{align}
\end{subequations}
summing over the two signs in the latter case.
Note that
$\detTwo(L)$ and $L^{-1}$ are computed in the 2-dimensional subspace spanned by $L$,
while $\hat{\bm\phi}$ is the unit vector orthogonal to this subspace. 

\sec{Differential Observables}
Using the waveform $f_{ij}$ one may compute differential
observables related to the energy.
Here we focus on the radiated energy in the rest
frame of black hole 1: $v_1\!\cdot\!P_\text{rad}$,
where $P^\mu_\text{rad}$ was given in Eq.~\eqref{eq:momdef}.
To compute the total energy radiated per unit solid angle we
leave out the angular $\d\sigma$ integration:
\begin{align}
  \frac{\d E_{\text{rad}}}{\d\sigma}
  =
  \frac1{32\pi G}
  \int\!\d u\, [\dot{f}_{ij}]^{2}
  \ .
\end{align}
In the low-velocity $v\ll 1$ expansion (and to leading order in $G$) we find:
\begin{widetext}
\begin{align}
\begin{aligned}
  \frac{\d E_{\text{rad}}}{\d\sigma} &=
  \frac{G^3m_{1}^{2}m_{2}^{2} v}{512 b^{3}}\Bigl [
  45(\cos^2\!\theta\cos^2\!\phi+\sin^2\!\phi)^2+
  109\sin^4\!\theta+630\sin^2\!\theta\sin^2\!\phi+
  354\sin^2\!\theta\cos^2\!\theta\cos^2\!\phi \\
  &\qquad\qquad\qquad
  +\frac{v}{2}\cos\theta\Bigl(
    135+30\sin^2\!\theta(44+61\sin^2\!\phi)+
    7\sin^4\!\theta(-200+264\sin^2\!\phi+45\sin^4\!\phi)\Bigr )
  \\
  &\qquad\qquad\qquad
  +
  \frac{v^{2}}{32}
  \Big(
    7545+
    9\sin^2\!\theta(5714+6627\sin^2\!\phi)+
    \sin^4\!\theta(-128104 + 27084\sin^2\!\phi+24255\sin^4\!\phi)
  \\
  &\qquad\qquad\qquad\qquad  +
    \sin^6\!\theta(53200 - 70728\sin^2\!\phi - 11790\sin^4\!\phi + 525\sin^6\!\phi)
  \Big)
    +\mathcal{O}(v^{3})\Bigr ]  +\mathcal{O}(G^4)\,.
\end{aligned}
\end{align}
\end{widetext}
This goes beyond the results of Kovacs and Thorne \cite{Kovacs:1978eu}
which stated only the leading $\mathcal{O}(v)$ post-Newtonian term of the
first line (and with which we agree).
We can straightforwardly extend this result to higher orders in $v$;
integrating over the two-sphere $\d\sigma$ yields the total radiated energy:
\begin{equation}
\frac{E_{\rm rad}}{\pi}=\frac{G^3m_1^2m_2^2v}{b^3}\left(
\frac{37}{15}+\frac{2393}{840}v^2+\frac{61703}{10080}v^4\right)+\cO(v^7)\,,
\end{equation}
which agrees with Ref.~\cite{Herrmann:2021lqe} to $\mathcal{O}(v^5)$.

\begin{figure}[t!]
\includegraphics[scale=0.4]{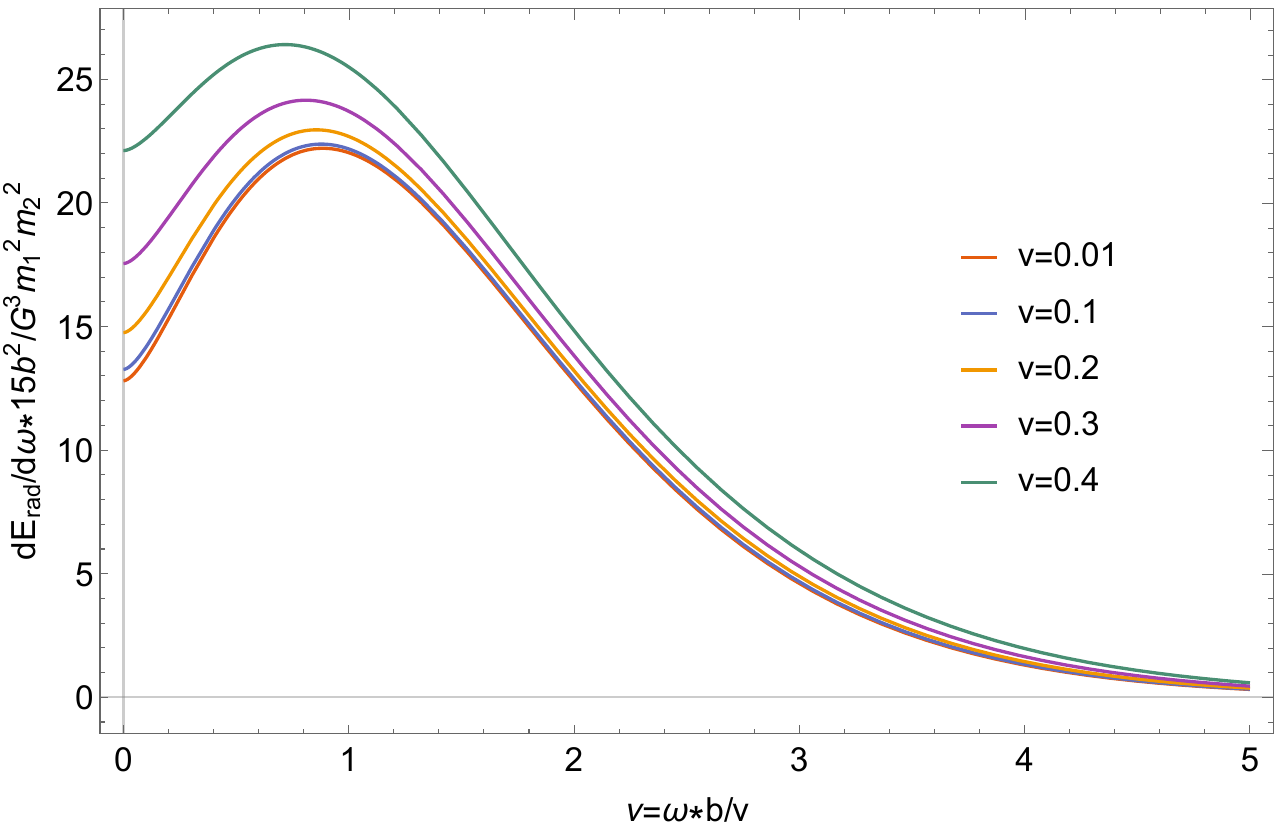}
\caption{The differential power spectrum (total energy radidated per unit frequency) given in Eq.~\eqn{psfinal} at $\cO(G^3)$, including terms up to order $v^{4}$ in a low-velocity expansion.}\label{fig:spectrum}
   \end{figure}

We can also compute the differential power spectrum
(total energy radiated per unit frequency):
\begin{align}
  \frac{\d E_{\text{rad}}}{\d\omega}
  =
  \frac{1}{16\pi G}
  \int\!\d \sigma
  |\omega f_{ij}(\omega) |^{2}
  \ ,
\end{align}
where $\omega$ is positive and $f_{ij}(\omega)$ is
the frequency-domain waveform:
\begin{align}
  f_{ij}(\omega) =
  \int\!\d u\ 
e^{i\omega u}f_{ij}(u)
  \ .
\end{align}
Performing the relevant integrals (with $\nu=\frac{b}{v}\omega>0$)
in the low-velocity expansion yields
\begin{widetext}
\begin{align}
\begin{aligned}
  \frac{\d E_{\text{rad}}}{\addcontentsline{}{}{}\d\omega}
  &=
  \frac{16G^3 m_1^2 m_2^2 \nu^2}{15b^2}
  \Big[
    12 
  \Big(
  (\sfrac{1}{3} + \nu ^2) {K_0}^2+3 \nu  K_0 K_1+ \left(1+\nu ^2\right) {K_1}^2
  \Big)
  \\
  &\qquad\qquad\qquad\qquad
  +
  \frac{v^2}{7}
  \Big(
  2 \left(5 \nu ^4-20 \nu ^2-64\right) {K_0}^2+4 \nu  \left(19 \nu ^2-8\right) K_0 K_1+\left(10 \nu ^4+3 \nu ^2+300\right) {K_1}^2
  \Big)
  \\
  &\qquad\qquad\qquad\qquad
  +
  \frac{v^4}{189}
  \Big(
  \left(20 \nu ^6+301 \nu ^4-1026 \nu ^2+1584\right) {K_0}^2+\nu  \left(120 \nu ^4-1523 \nu ^2+4518\right) K_0 K_1
  \\
  &\qquad\qquad\qquad\qquad\qquad
  +\left(20 \nu ^6+371 \nu ^4+4417 \nu ^2+13860\right) {K_1}^2
  \Big)+\cO(v^5)\Big]
  +{\mathcal O}(G^4)\,,  \label{psfinal}
\end{aligned}
\end{align}
where $K_0=K_0(\nu)$ and $K_1=K_1(\nu)$ are modified Bessel functions of the second kind.
This series can also be generated to any desired order in $v$,
and in Fig.~\ref{fig:spectrum} we plot the power spectrum for a selection
of different velocities.
Again, the first line at leading order in $v$ agrees with Ref.~\cite{Kovacs:1978eu}.

\sec{Visualizations}
Sample visualizations of the complete waveforms as they evolve
with retarded time $u$ are available for 
\begin{itemize}
\item $f_{+}(u,\theta,\phi)|_{v=0.2}$ at 
  \href{https://box.hu-berlin.de/f/73b510f4dae1408ba612/}{\tt https://box.hu-berlin.de/f/73b510f4dae1408ba612/},
  \item $f_{\times}(u,\theta,\phi)|_{v=0.2}$ at 
  \href{https://box.hu-berlin.de/f/6c1ff4812f7448ca913d/}{\tt https://box.hu-berlin.de/f/6c1ff4812f7448ca913d/}.
\end{itemize}
\end{widetext}

\bibliography{paper_wqft_radiation}

\end{document}